\documentclass[twocolumn,appendixfloats,twocolappendix]{aastex6}
\usepackage{mathtools}
\usepackage{graphicx}
\usepackage{physics}
\usepackage{xcolor}
\usepackage{hyperref}

\newcommand{\HI}{H~\textsc{i}}

\newcommand{\HeII}{He~\textsc{ii}}
\newcommand{\HeIII}{He~\textsc{iii}}

\begin{document}

\shorttitle{Helium Reionization Simulations III}
\shortauthors{La Plante et al.}

\title{Helium Reionization Simulations. III. The Helium Lyman-$\alpha$ forest}
\author{Paul La Plante\altaffilmark{1,2},
  Hy Trac\altaffilmark{2},
  Rupert Croft\altaffilmark{2},
  and Renyue Cen\altaffilmark{3}}
\altaffiltext{1}{Center for Particle Cosmology, Department of Physics and
  Astronomy, University of Pennsylvania, Philadelphia, PA 19104, USA, \\ \href{mailto:plaplant@sas.upenn.edu}{plaplant@sas.upenn.edu}}
\altaffiltext{2}{McWilliams Center for Cosmology, Department of Physics,
  Carnegie Mellon University, Pittsburgh, PA 15213, USA}
\altaffiltext{3}{Department of Astrophysical Science, Princeton University,
  Princeton NJ 08544, USA}

\begin{abstract}
  In \citet{laplante_etal2017}, we presented a new suite of hydrodynamic
  simulations with the aim of accurately capturing the process of helium
  \textsc{ii} reionization. In this paper, we discuss the observational
  signatures present in the \HeII\ Ly$\alpha$ forest. We show that the effective
  optical depth of the volume $\tau_\mathrm{eff}$ is not sufficient for
  capturing the ionization state of helium~\textsc{ii}, due to the large
  variance inherent in sightlines. However, the \HeII\ flux PDF can be used to
  determine the timing of helium~\textsc{ii} reionization. The amplitude of the
  one-dimensional flux power spectrum can also determine the ionization state of
  helium~\textsc{ii}. We show that even given the currently limited number of
  observations ($\sim$50 sightlines), measurements of the flux PDF can yield
  information about helium~\textsc{ii} reionization. Further, measurements using
  the one-dimensional power spectrum can provide clear indications of the timing
  of reionization, as well as the relative bias of sources of ionizing
  radiation.
\end{abstract}

\keywords{cosmology: theory --- intergalactic medium --- large-scale
  structure of the universe --- methods: numerical --- quasars:
  general}

\section{Introduction}
\label{sec:intro}
There has been much interest in understanding the reionization of
helium~\textsc{ii}, using semi-analytic methods
\citep{gleser_etal2005,furlanetto_oh2008a,furlanetto_oh2008b,furlanetto_oh2009,dixon_etal2014},
numerical simulations
\citep{croft_etal1997,mcquinn_etal2009,mcquinn_etal2011,compostella_etal2013,compostella_etal2014,puchwein_etal2015,bolton_etal2016},
and observations
\citep{jakobsen_etal1994,reimers_etal1997,zheng_etal2008,dixon_furlanetto2009,syphers_shull2014,worseck_etal2014}. Helium~\textsc{ii}
reionization is thought to be driven by highly energetic photons emitted by
quasars. Due to photoheating of gas in the intergalactic medium (IGM) from these
high-energy photons, helium~\textsc{ii} reionization leaves an important
signature on the thermal state of the IGM. Knowledge of the thermal state is
important for making measurements of quantities related to the Ly$\alpha$
forest, such as the free-streaming length of warm dark matter
\citep{viel_etal2005,irsic_etal2017}. However, such temperature measurements are
difficult to make and have large systematic or statistical uncertainties
\citep{schaye_etal1999,mcdonald_etal2001,becker_etal2011a,boera_etal2014}. Further,
these methods rely on correctly calibrating the state of the hydrogen Ly$\alpha$
forest with the gas temperature, which is fraught with difficulty.

A more appealing approach is to measure the ionization state of helium directly,
without relying on calibration. Just as the Ly$\alpha$ transition for neutral
hydrogen (\HI) appears as absorption features in spectra of radiation from
distant quasars, so too does the transition for singly ionized helium (\HeII)
appear. This feature appears at 304 \AA\ in the rest-frame of the absorbing gas,
a shift of a factor of four in frequency space compared to the hydrogen
transition due to the additional proton in the helium nucleus. As with the \HI\
Ly$\alpha$ forest, the very high transition strength means a very small amount
of singly ionized helium can lead to total absorption of the incoming
radiation. Typically, neutral fractions of $f_\mathrm{HeII} \gtrsim 10^{-3}$ can
produce a Gunn-Peterson trough \citep{gunn_peterson1965}, making detection of
the early stages of helium~\textsc{ii} reionization difficult. Despite this
difficulty, measuring the ionization status of helium from the \HeII\ Ly$\alpha$
forest provides a more direct probe than using temperature measurements or the
\HI\ Ly$\alpha$ forest.

Part of the difficulty in observing the \HeII\ Ly$\alpha$ forest lies in
contamination of high-density systems at lower redshift. Lyman-limit systems
(LLSs) and damped Ly$\alpha$ systems (DLAs) which are at intermediate redshift
(say at $z_\mathrm{LLS}$) between the comparatively high-redshift IGM gas we are
interested in observing (say at $z_\mathrm{IGM}$) and observers on Earth can
absorb much of the radiation above the ionization potential of hydrogen at 912
\AA. Quantitatively, if $912(1+z_\mathrm{LLS}) \gtrsim 304(1+z_\mathrm{IGM})$,
then the lower-redshift LLS or DLA will obfuscate the \HeII\ Ly$\alpha$ forest
of interest. Due to the relative abundance of LLSs and DLAs at low redshift,
only a small number of quasar sightlines are suitable for measuring the \HeII\
Ly$\alpha$ forest \citep{moller_jakobsen1990,zheng_etal2005}. Indeed, despite
having more than 150,000 quasar sightlines from BOSS alone
\citep{dawson_etal2013}, to date there have been only about 50 sightlines for
which the \HeII\ Ly$\alpha$ forest has been measured
\citep{syphers_etal2009a,syphers_etal2009b,syphers_etal2012}. \footnote{It
  should be noted, though, that a single sightline of, \textit{e.g.}, 100
  physical Mpc, can yield multiple measurements of a given statistic by dividing
  the total sightline into multiple smaller segments of moderate size
  (\textit{e.g.}, 10 physical Mpc as in \citealt{worseck_etal2014}).}  These
measurements have provided significant insight to the general picture of
helium~\textsc{ii} reionization: at redshifts $z > 3$, a Gunn-Peterson trough
has been detected \citep{jakobsen_etal1994,zheng_etal2008,syphers_shull2014};
below this redshift, helium~\textsc{ii} reionization becomes patchy, showing
extended regions of absorption and transmission corresponding to the ionization
level of the gas \citep{reimers_etal1997}; finally, by redshift $z \sim 2.7$,
helium appears to be totally reionized
\citep{dixon_furlanetto2009,worseck_etal2011}. However, information beyond this
general picture is difficult to glean from the current limited set of \HeII\
spectra. To this end, measurements providing additional information about
helium~\textsc{ii} reionization is an important application of current and
ongoing research.

In \citet{laplante_trac2016} (hereafter Paper~I) of this paper series, we
provided a method by which simulation volumes can be populated with quasars in
order to reproduce the quasar luminosity function (QLF) at various redshift
epochs \citep{masters_etal2012,ross_etal2013,mcgreer_etal2013} as well as quasar
clustering \citep{white_etal2012}. In \citet{laplante_etal2017} (hereafter
Paper~II), we presented a new suite of large-scale simulations with the purpose
of exploring helium~\textsc{ii} reionization. These simulations include
$N$-body, hydrodynamics, and radiative transfer solved simultaneously, which
allows us to capture the evolution of the IGM with newfound accuracy. Based on
the output of these simulations, we are able to generate synthetic Ly$\alpha$
sightlines for \HI\ and \HeII. In this paper, we present specific results about
the \HeII\ spectra, and discuss ways to learn about the timing of
helium~\textsc{ii} reionization.

We organize the rest of this paper as follows. In Section~\ref{sec:radhydro}, we
briefly discuss our suite of simulations. In Section~\ref{sec:lya}, we discuss the
\HeII\ Ly$\alpha$ forest, and various measurements that can be made using
the spectra. In Section~\ref{sec:discussion}, we discuss prospects for detecting
helium~\textsc{ii} reionization properties given the current measurements. In
Section~\ref{sec:conclusion}, we summarize our findings. Throughout this work, we
assume a $\Lambda$CDM cosmology with $\Omega_m = 0.27$, $\Omega_\Lambda = 0.73$,
$\Omega_b = 0.045$, $h = 0.7$, $\sigma_8 = 0.8$, and $Y_\mathrm{He} =
0.24$. These values are consistent with the \textit{WMAP}-9 year results
\citep{hinshaw_etal2013}.

\section{Radiation-Hydrodynamic Simulations}
\label{sec:radhydro}

In Paper~II, we presented a new suite of hydrodynamic simulations with radiative
transfer, conducted with the goal of studying helium~\textsc{ii}
reionization. Here, we summarize the properties of the simulations that are
relevant to this paper's results. Radiation-hydrodynamic simulations are run
with the RadHydro code, which combines $N$-body and hydrodynamic algorithms
\citep{trac_pen2004} with an adaptive ray-tracing algorithm \citep{trac_cen2007}
to directly and simultaneously solve collisionless dark matter, collisional gas
dynamics, and radiative transfer of ionizing photons. The simulations in
Paper~II employ 2048$^3$ dark matter particles, and 2048$^3$ hydrodynamic
resolution elements in a fixed-grid Eulerian scheme.  The grid for radiative
transfer contains 512$^3$ resolution elements. The simulation code has already
been used to study hydrogen reionization
\citep{trac_cen2007,trac_etal2008,battaglia_etal2013a}. For additional details
about the simulations, see Paper~II.

The simulations contain two features in particular that bear mentioning. First,
the simulations include a patchy model for hydrogen reionization developed in
\citet{battaglia_etal2013a}. The midpoint of reionization has been set such that
$\bar{z}_\mathrm{re} = 8$, but in general, regions of high-density undergo
reionization before regions of low density. By incorporating an ``inside-out''
reionization scenario, we ensure that the thermal state of the IGM before
helium~\textsc{ii} reionization accurately reflects the impact of hydrogen
reionization. Second, the contribution of galaxies to the UV background
$\Gamma_\mathrm{gal}$ is modified on-the-fly in order to reproduce the observed
effective optical depth $\tau_\mathrm{eff}$ of the \HI\ Ly$\alpha$ forest. The
quantity $\tau_\mathrm{eff}$ is related to the flux of the Ly$\alpha$ forest
$F$, defined for every location in the volume as $F \equiv e^{-\tau}$. In this
expression, $\tau$ is the optical depth of Ly$\alpha$ radiation. Note that
$\bar{F} \neq e^{-\bar{\tau}}$. Values of $F \sim 0$ represent total absorption
(typically the result of a high density of neutral hydrogen), and values of
$F \sim 1$ represent total transmission. It is then possible to define the
effective optical depth of the entire volume, namely as:
\begin{equation}
\ev{F}_\mathrm{HI} = e^{-\tau_\mathrm{eff,HI}},
\label{eqn:tau}
\end{equation}
where $\ev{F}_\mathrm{HI}$ is the average flux of the \HI\ Ly$\alpha$ forest
of the volume, with an analogous definition for \HeII. Specifically, we match
$\tau_\mathrm{eff}$ as parameterized in \citet{lee_etal2015}. These results are
based primarily on data from the seventh data release of the Sloan Digital Sky
survey (SDSS DR7) presented in \citet{becker_etal2013}. Modifying
$\Gamma_\mathrm{gal}$ while the simulations are running means we do not need to
renormalize the Ly$\alpha$ forest in post-processing, as previous studies of
the Ly$\alpha$ forest have done (\textit{e.g.},
\citealt{bolton_etal2009a}). This feature allows us to more easily compare the
results between simulations and observations.

\begin{figure}[t]
  \centering
  \includegraphics[width=0.45\textwidth]{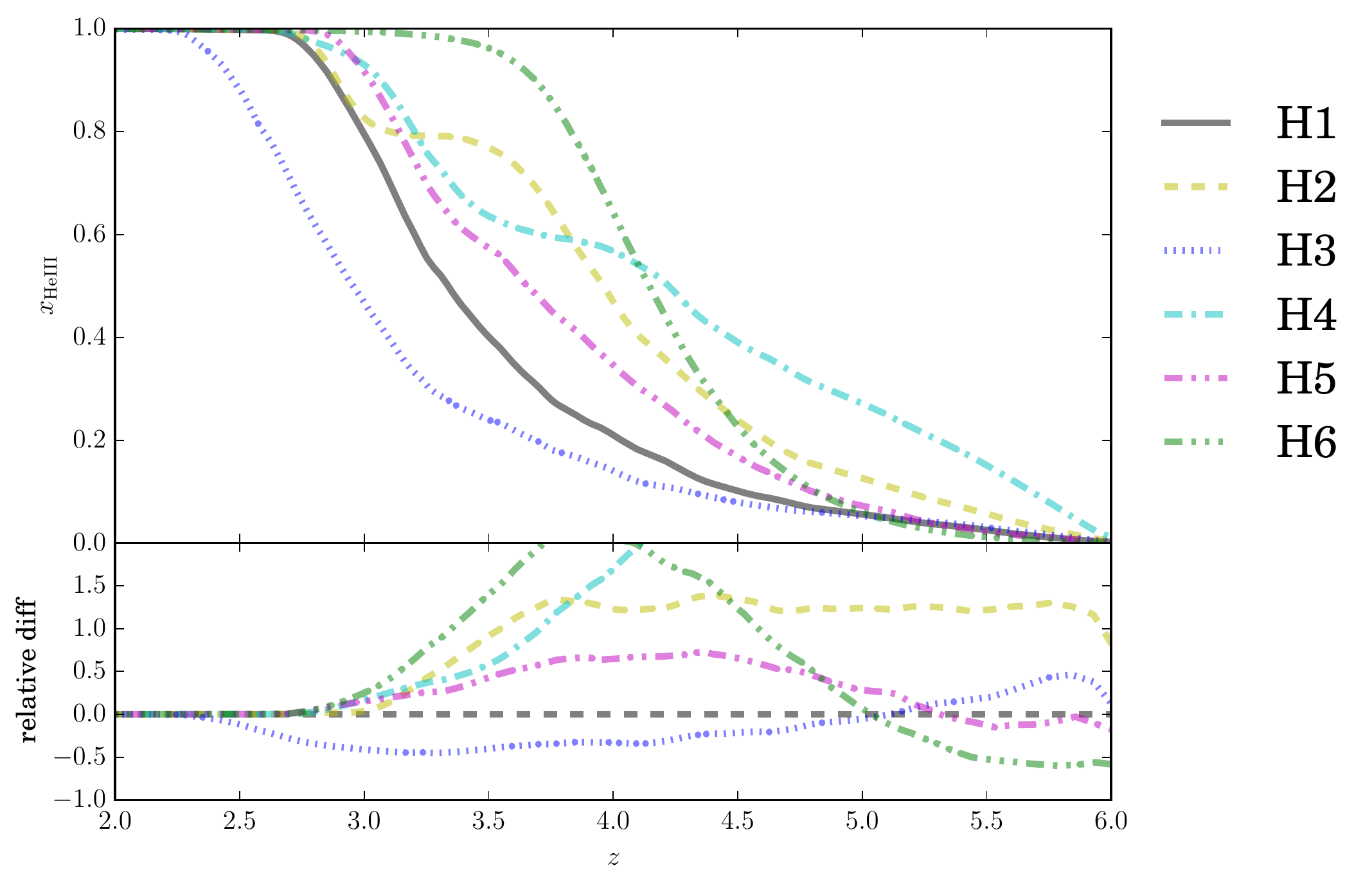}
  \caption{The ionization level of simulations as a function of redshift for the
    simulation suite presented in Paper~II. Simulation~H1 is the fiducial
    reionization scenario. Simulations~H2 and H3 increase and decrease the
    amplitude of the QLF, respectively, by a factor of two. Simulation~H4 uses
    the same sources as H1, but increases the photon production per quasar by a
    factor of two. Simulation~H5 uses a slightly different QLF from the other
    simulations.  Simulation~H6 uses a uniform UV-background instead of explicit
    quasar sources. The models are discussed further in
    Section~\ref{sec:radhydro}.}
  \label{fig:fHeIII}
\end{figure}

As explained in detail in Paper~I, the simulation volumes are populated with
quasars such that the observed QLF is matched between redshifts
$2 \leq z \leq 6$ \citep{masters_etal2012,ross_etal2013,mcgreer_etal2013}, as
well as the clustering measurements at $z \sim 2.4$ \citep{white_etal2012}. For
individual quasar objects, we use the SED from \citet{lusso_etal2015}, which has
a spectral index of $\alpha = 1.7$ ($f_\nu \propto \nu^{-\alpha}$) for
$\lambda \leq 912$ \AA, and a spectral index of $\alpha = 0.61$ for
$\lambda > 912$ \AA.

In Paper~II, we presented a suite of six simulations, with different quasar
properties. We will now briefly summarize each of these simulations. Simulation
H1 is the fiducial reionization model, which uses the QLF combining the various
measurements at distinct redshift epochs and the SED of
\citet{lusso_etal2015}. Simulation~H2 increases the amplitude of the QLF by a
factor of 2, leading to an earlier reionization scenario. Simulation~H3
decreases the amplitude of the QLF by a factor of 2, leading to a late
reionization time. Simulation~H4 increases the normalization of the SED by a
factor of 2, so that a given quasar with a given magnitude $M$ will have a
luminosity at 912 \AA\ $L_{912}$ that is two times greater as that provided by
the SED of \citet{lusso_etal2015}. Simulation~H5 uses a slightly different
method for combining the QLF from distinct redshift epochs than Simulation~H1,
but uses the same SED. Simulation~H6 does not have explicit quasar sources, but
instead uses a uniform UV background with the photoionization and photoheating
rates as specified by \citet{haardt_madau2012}. Rather than simply using the
rates ``as is,'' we scale them to match the observed value of
$\tau_\mathrm{eff,HI}$, as with the other simulations. See Paper~II for further
details about each of the simulations. Figure~\ref{fig:fHeIII} shows the
ionization fraction of the various simulations as a function of redshift.

\section{He~\textsc{ii} Ly$\alpha$ Forest Measurements}
\label{sec:lya}

At each time step in the simulation, we generate synthetic Ly$\alpha$
sightlines on-the-fly for \HI\ and \HeII. The measurements of
$\tau_\mathrm{eff,HI}$ for the \HI\ sightlines allow for modifying
$\Gamma_\mathrm{gal}$ to ensure that the value is matched at all times in the
simulation. Determining \HeII\ from the synthetic sightlines allows for a more
straightforward connection with the ionization state of helium in the volume.
We will now turn to specific observables related to the \HeII\ Ly$\alpha$
forest.

\subsection{Effective Optical Depth}
\label{sec:taueff}
The effective optical depth $\tau_\mathrm{eff}$, as noted in
Eqn.~(\ref{eqn:tau}), is defined in terms of the average flux in the volume. As
with the \HI\ Ly$\alpha$ transition, the strength of the \HeII\ transition
ensures that only a very small amount of singly-ionized helium is necessary to
completely absorb incoming radiation. As a result, measurement of
$\tau_\mathrm{eff,HeII}$ is most sensitive to the end of reionization. Further,
due to the very large comoving size of \HeIII\ regions (typically tens of Mpc in
diameter), there is a large variation between different sightlines in the
simulation volume, or even along a given sightline. One significant reason for
this is that the correlation length $s_0$ of quasars, defined by the
three-dimensional two-point correlation function $\xi(s) = (s/s_0)^{-2}$
(\textit{e.g.}, \citealt{white_etal2012}), is comparable to, but slightly larger
than, the mean free path of $h\nu = 54.4$ eV photons for
$f_\mathrm{HeIII} \sim 0.8$ at $z \sim 3$. Accordingly, until there is overlap
of ionized regions, there are large contiguous regions of \HeII\ and
\HeIII. Ultimately, this results in large variation along a line of
sight. Furthermore, these ionization regions typically do not overlap until the
tail-end of reionization. This variation is especially prevalent while
helium~\textsc{ii} reionization is proceeding. In other words, due to the large
coherence of the doubly ionized regions, the observed optical depth can vary
greatly from sightline to sightline, and so there should be a large variance in
the measurements. This variation is in addition to any inherent variance in
$\tau_\mathrm{eff,HeII}$, primarily due to density fluctuations.

\begin{figure}[t]
  \centering
  \includegraphics[width=0.45\textwidth]{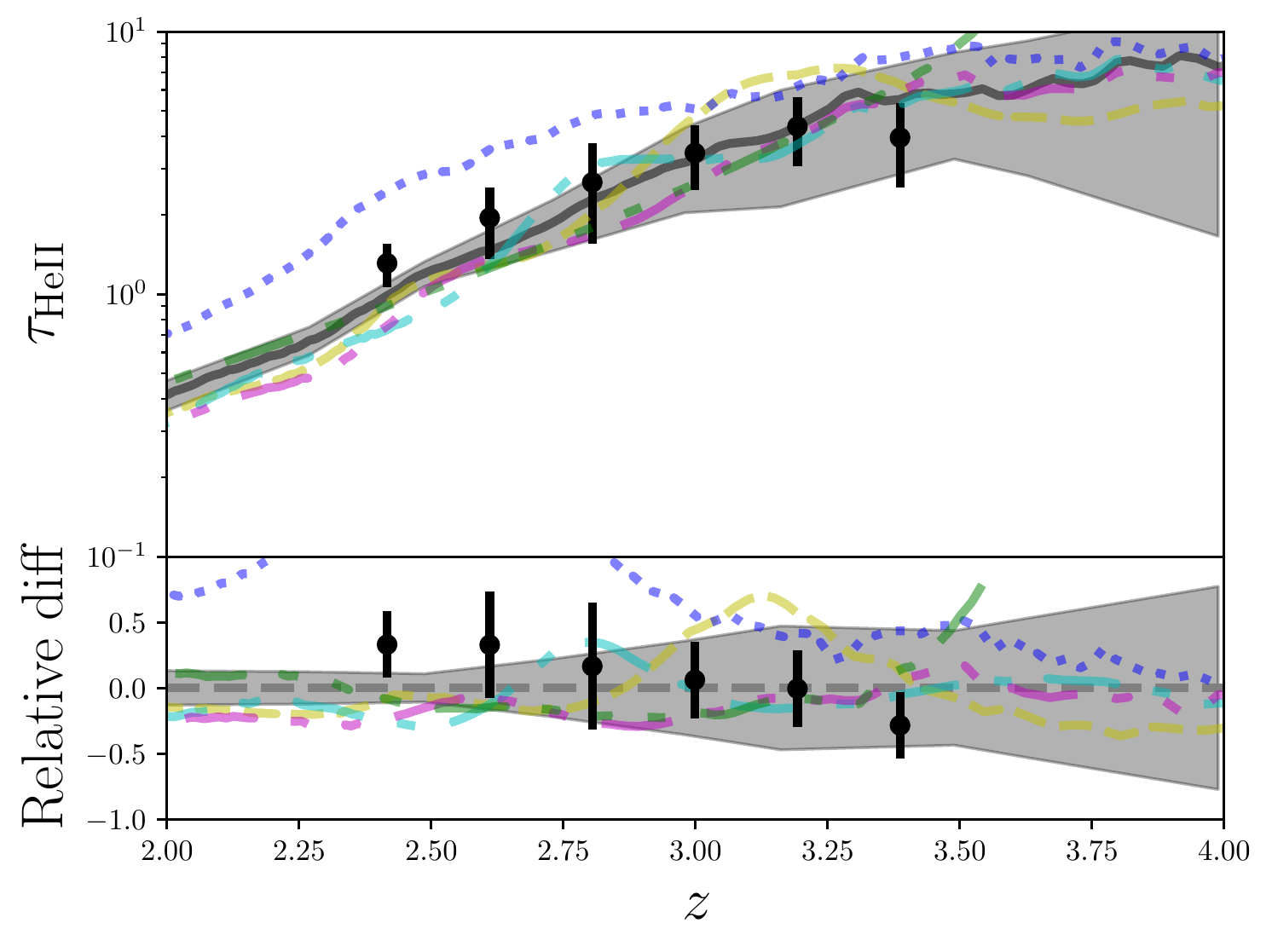}
  \caption{The effective optical depth of singly ionized helium
    $\tau_\mathrm{eff,HeII}$ as a function of redshift for the suite of
    simulations. The black dots represent the binned observational data from
    \citet{worseck_etal2014}, which are taken from HST/COS data. Color and line
    styles for the simulations are the same as in Figure~\ref{fig:fHeIII}. The
    top panel shows the results for the simulations presented in Paper~II, and
    the bottom panel shows the relative difference to the fiducial
    simulation. As can be seen, there is a large degree of scatter in the
    measurements. By extension, none of the simulations is clearly
    disfavored. See Section~\ref{sec:taueff} for additional discussion.}
  \label{fig:tauHeII}
\end{figure}

Figure~\ref{fig:tauHeII} shows $\tau_\mathrm{eff,HeII}$ as a function of
redshift averaged over the whole simulation volume. The Figure also includes
observational data from \citet{worseck_etal2014}. These quasar spectra were
taken using the cosmic origins spectrograph (COS) on the Hubble Space Telescope
(HST). These spectra measure $\tau_\mathrm{eff,HeII}$ for segments of about 10
proper Mpc. The large sightline-to-sightline variation is evident in the
observational data, which show very different values of $\tau_\mathrm{eff,HeII}$
at the same redshift. The results from most of the simulations are largely
consistent with the data at the redshifts for which data is available
($2.5 \lesssim z \lesssim 3.5$). The main exception to this is Simulation~H3,
which completes reionization at a significantly later time than the other
simulations. Quantitatively, the redshift when the volume of Simulation~H3
reaches an ionization fraction
$x_\mathrm{HeIII} \equiv n_\mathrm{HeIII}/n_\mathrm{He}$ of 99\% is
$z_{99} \sim 2.23$, compared to the timing of reionization in the fiducial
scenario of $z_{99} \sim 2.69$. Given that this simulation completes
reionization significantly later than the other ones, it is not surprising that
the value of $\tau_\mathrm{eff,HeII}$ in Simulation~H3 is significantly higher
than that of the other ones.

\begin{figure*}[t]
  \centering
  \resizebox{0.48\textwidth}{!}{\includegraphics{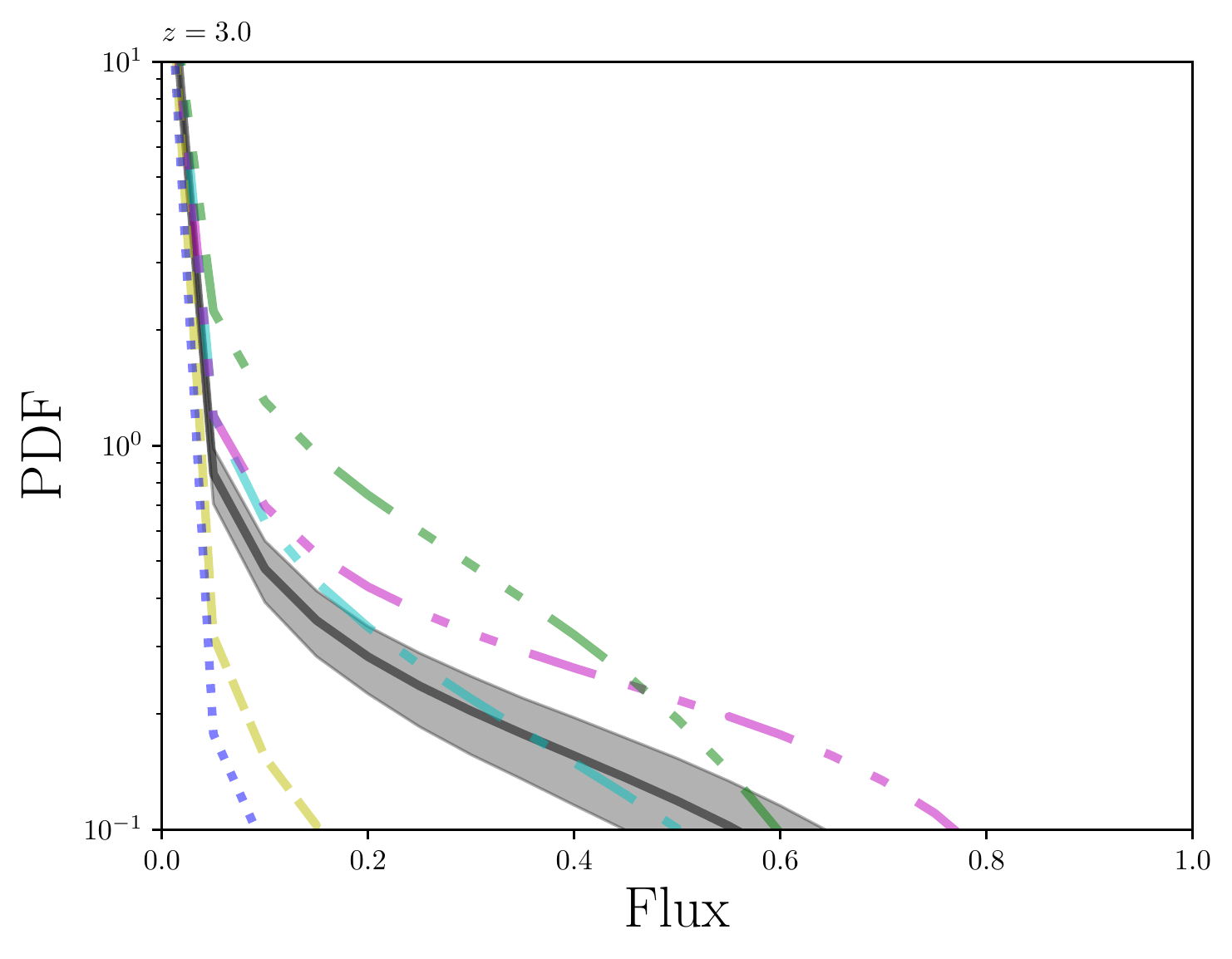}}%
  \resizebox{0.48\textwidth}{!}{\includegraphics{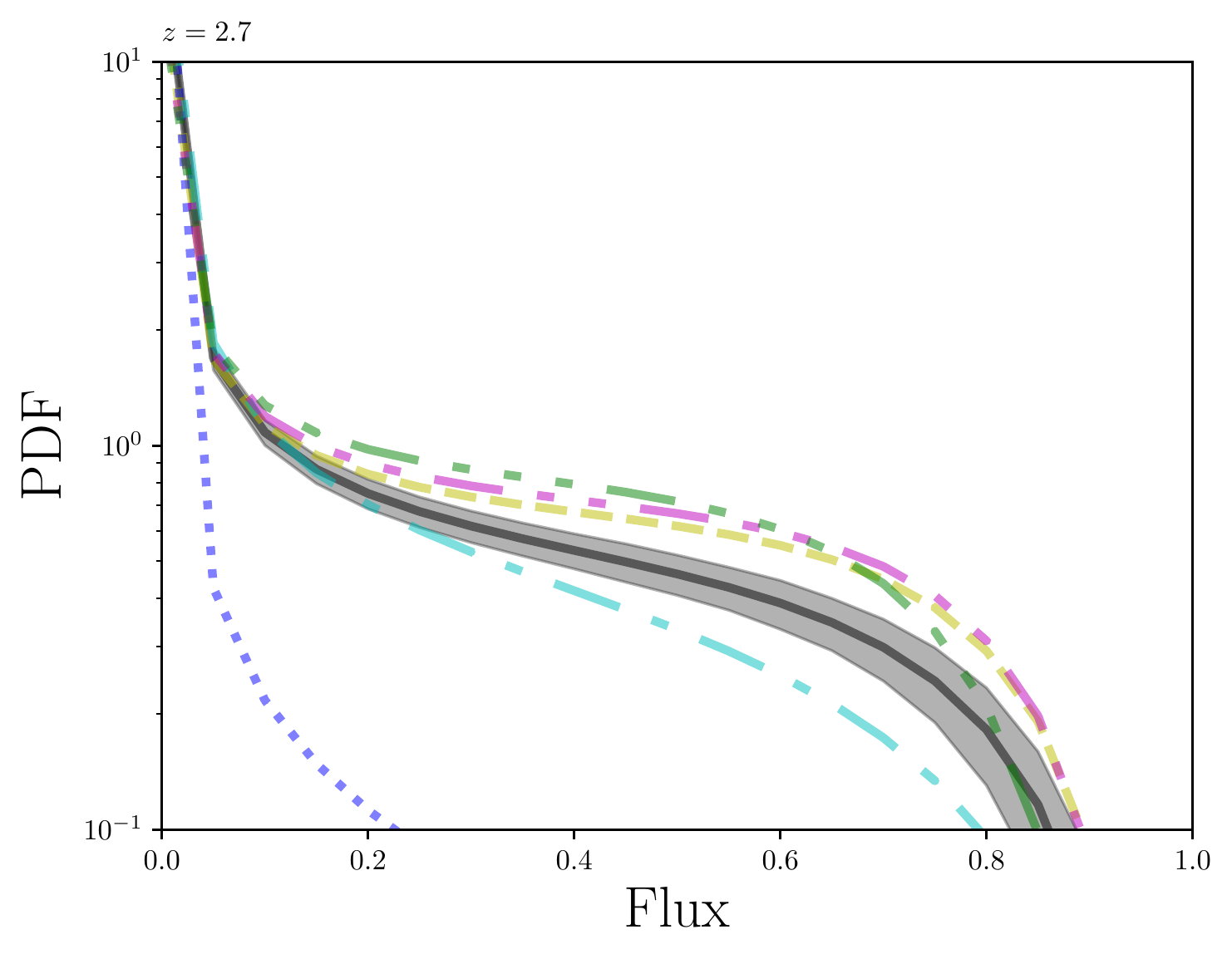}}\\
  \resizebox{0.48\textwidth}{!}{\includegraphics{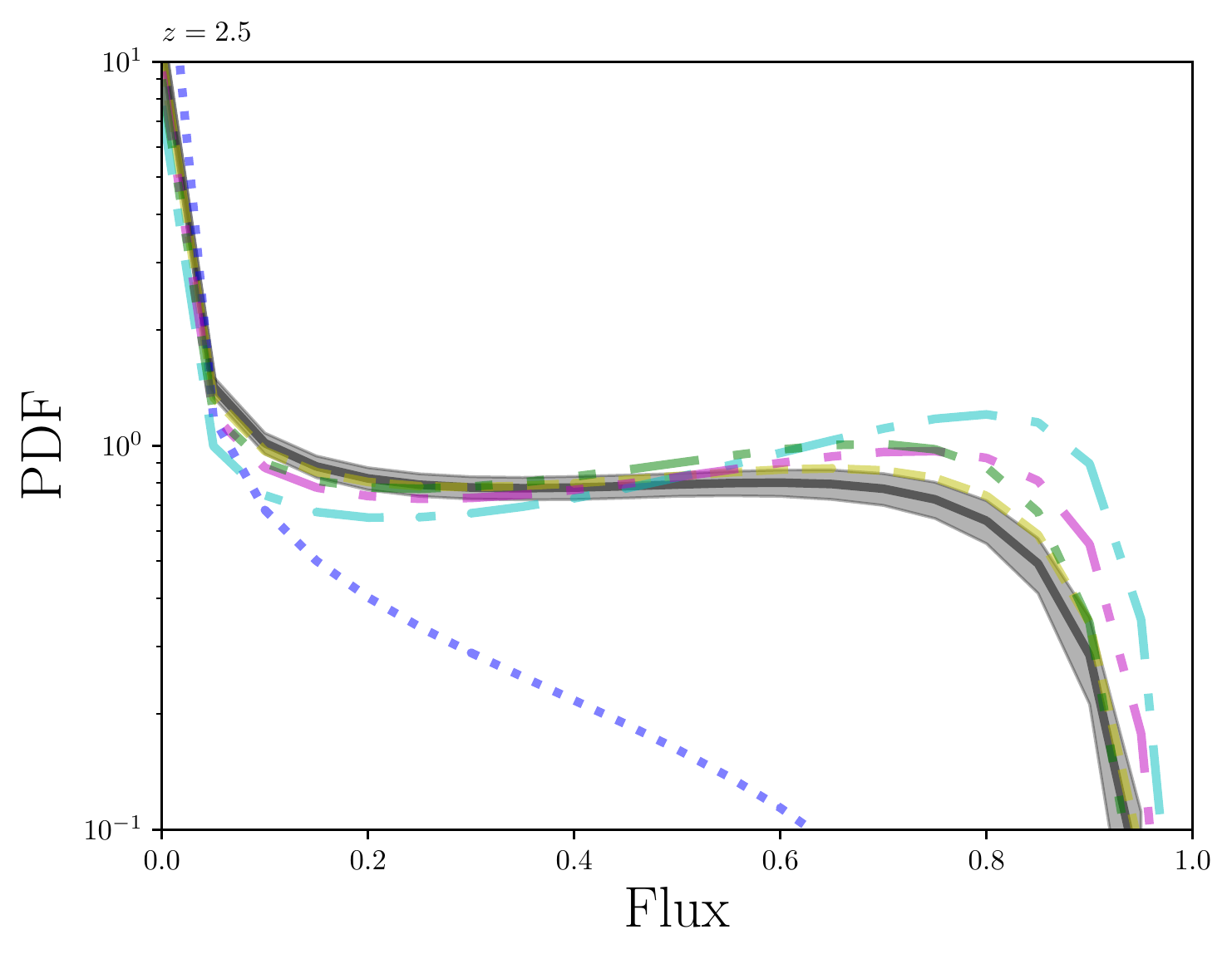}}%
  \resizebox{0.48\textwidth}{!}{\includegraphics{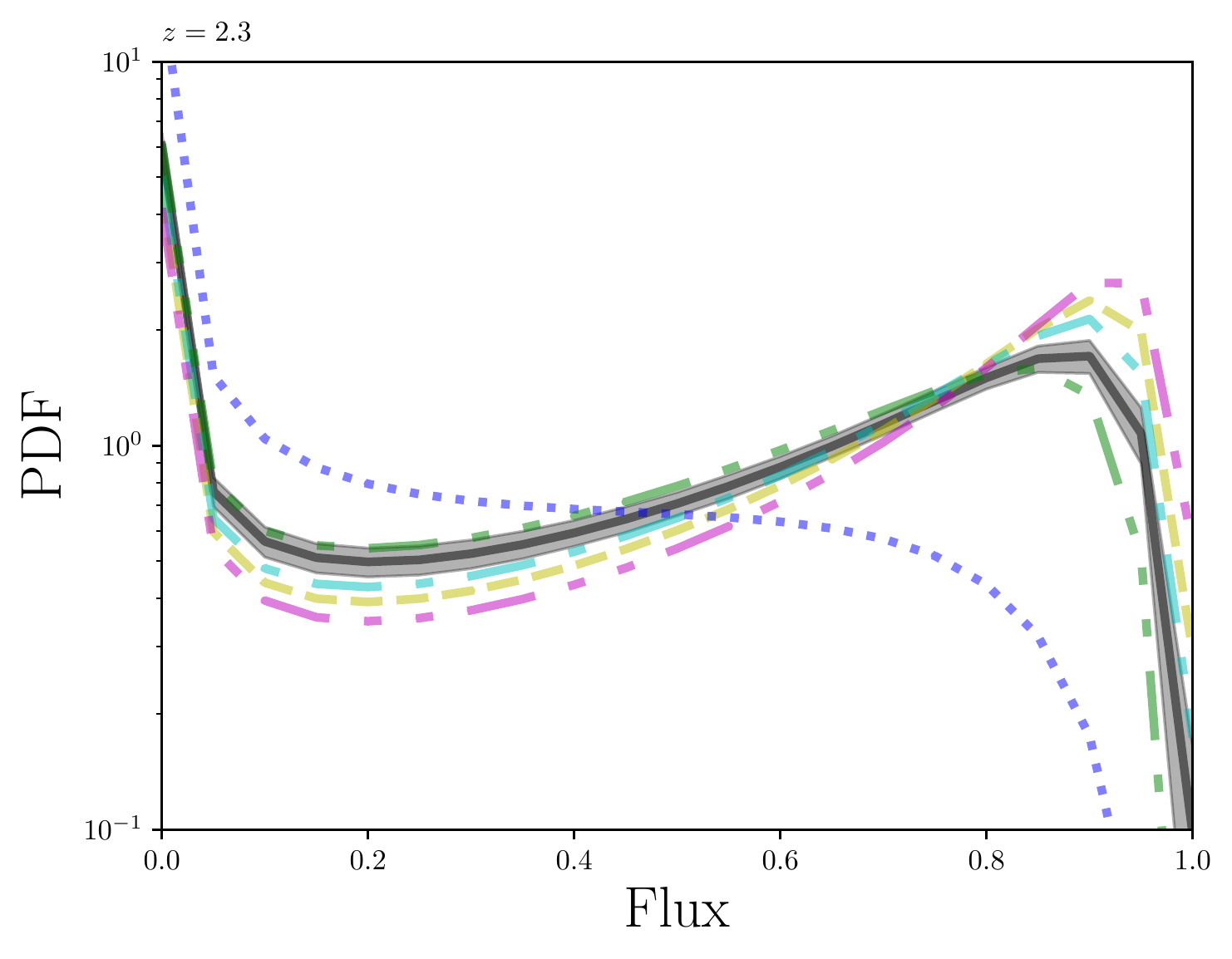}}\\
  \caption{The flux PDF of the HeII Ly$\alpha$ forest at redshifts $z \sim 3$
    (top left), $z \sim 2.7$ (top right), $z \sim 2.5$ (bottom left), and
    $z \sim 2.3$ (bottom right). The flux PDF is very sensitive to the tail end
    of reionization: most of the simulations have an ionization fraction
    $x_\mathrm{HeIII} \gtrsim 0.99$ at $z \sim 2.7$, and yet have a
    comparatively low number of pixes with high transmission ($F \gtrsim
    0.5$). Nevertheless, the ionization fraction can be determined from the
    overall shape of the PDF: the flux PDF of Simulation~H3 at redshift
    $z \sim 2.3$ looks comparable to the other simulations at earlier times,
    such as at redshift $z \sim 2.7$. Accordingly, the flux PDF has a similar
    shape at comparable helium~\textsc{iii} ionization fractions, which will be
    discussed more in Figure~\ref{fig:xHeII_pdf}. The shaded regions show the
    error in the determination of the mean of Simulation~HI computed using
    bootstrap resampling. For more details, see the discussion in
    Section~\ref{sec:fluxpdf}.}
  \label{fig:fluxPDF}
\end{figure*}

The gray shaded regions in Figure~\ref{fig:tauHeII} show the 1$\sigma$ standard
deviation estimated by calculating the standard deviation of flux
$\sigma_F = \ev{F^2} - \ev{F}^2$ computed using all sightlines in the
volume. This standard deviation $\sigma_F$ is then converted into the standard
deviation of optical depth $\sigma_\tau$ using standard error propagation
methods. Note that these methods implicitly assume that the distribution is
Gaussian. The values in flux are generally not Gaussian, with most of the values
tending to be either 0 or 1 (see Figure~\ref{fig:fluxPDF}). Thus, the shaded
regions of the plot are overly optimistic, and do not capture the true variance
of $\tau_\mathrm{eff,HeII}$ that is present in the volume. In particular, the
shaded regions should extend to much higher values of $\tau_\mathrm{eff,HeII}$,
since these represent sightlines with $\ev{F} \sim 0$.

An alternative to using error propagation is to use a different statistic for
capturing the distribution of $\tau_\mathrm{HeII}$, or to use different sampling
methods. For instance, instead of the mean and standard deviation, one could
calculate $\ev{F} = \tau_\mathrm{eff,HeII}$ for each sightline, and then
calculate the median and central $\sim$68\% of central values (corresponding to
1$\sigma$). On the other hand, this approach poses problems related to the large
difference of the median and the mean of the distribution of
$\tau_\mathrm{eff,HeII}$. In particular, at moderate redshift ($z \gtrsim 3$),
the median of $\tau_\mathrm{eff,HeII}$ measured per-sightline is significantly
larger than the mean. As discussed above, for ionization fractions
$x_\mathrm{HeIII} \lesssim 99\%$, most sightlines demonstrate very high
absorption, and so $\ev{F} \approx 0$. Though these high-absorption sightlines
do not significantly alter $\ev{F}$ when computed for the entire volume, they
represent a significant number of individual sightlines, making the median
noticeably distinct from the mean.

Another possible way to circumvent the difficulties associated with error
propagation is to use bootstrap resampling to provide an estimate on the
distribution of the mean. The relevant parameter then becomes how many samples
to use when estimating the variation on the mean. As noted in
Section~\ref{sec:intro}, to date there have been roughly 50 \HeII\ sightlines
observed. Na\"{i}vely, one might assume that using 50 sightlines in the
bootstrap calculation would be a way to determine the variance of
$\tau_\mathrm{eff,HeII}$ that should be observed. Using 50 sightlines in the
bootstrap calculation does not accurately represent the large intrinsic scatter
seen in determining $\tau_\mathrm{eff,HeII}$ from different sightlines at a
given redshift. Furthermore, the variance estimated from 50 sightlines in a
bootstrap realization does not reflect the state of observations entirely, since
there is a relatively broad redshift distribution of the observed sightlines
(see the observed points in Figure~\ref{fig:tauHeII}). Furthermore, the variance
determined from using bootstrap resampling will only provide an estimate on the
error in determining the mean, rather than capturing the intrinsic scatter of
$\tau_\mathrm{eff,HeII}$.

The shaded regions still provide valuable information about our ability to
distinguish between various scenarios. In particular, note that at redshifts
$z \gtrsim 3$, the shaded regions encompass many of the observed points (though,
as mentioned above, the shaded error regions should extend to higher values of
$\tau_\mathrm{eff,HeII}$), as well as the values of $\tau_\mathrm{eff,HeII}$
from distinct scenarios. Thus, the quantity $\tau_\mathrm{eff,HeII}$ with only a
handful of measurements is not a reliable method for determining the history of
He~\textsc{ii} reionization.

\subsection{Flux PDF}
\label{sec:fluxpdf}
As discussed in Paper~II, another tool for analyzing the ionization state of the
medium is the flux PDF. This measurement captures the distribution of flux for
all pixels. As with $\tau_\mathrm{eff,HeII}$, this statistic is most sensitive
to the tail-end of reionization. Due to the low number of \HeII\ pixels with
high transmission ($F \gtrsim 0.5$) before the end of reionization
($x_\mathrm{HeIII} \gtrsim 0.99$), the flux PDF cannot provide detailed
information while reionization is underway. However, it can still provide
valuable information about the timing of reionization.

\begin{figure}[t]
  \centering
  \includegraphics[width=0.45\textwidth]{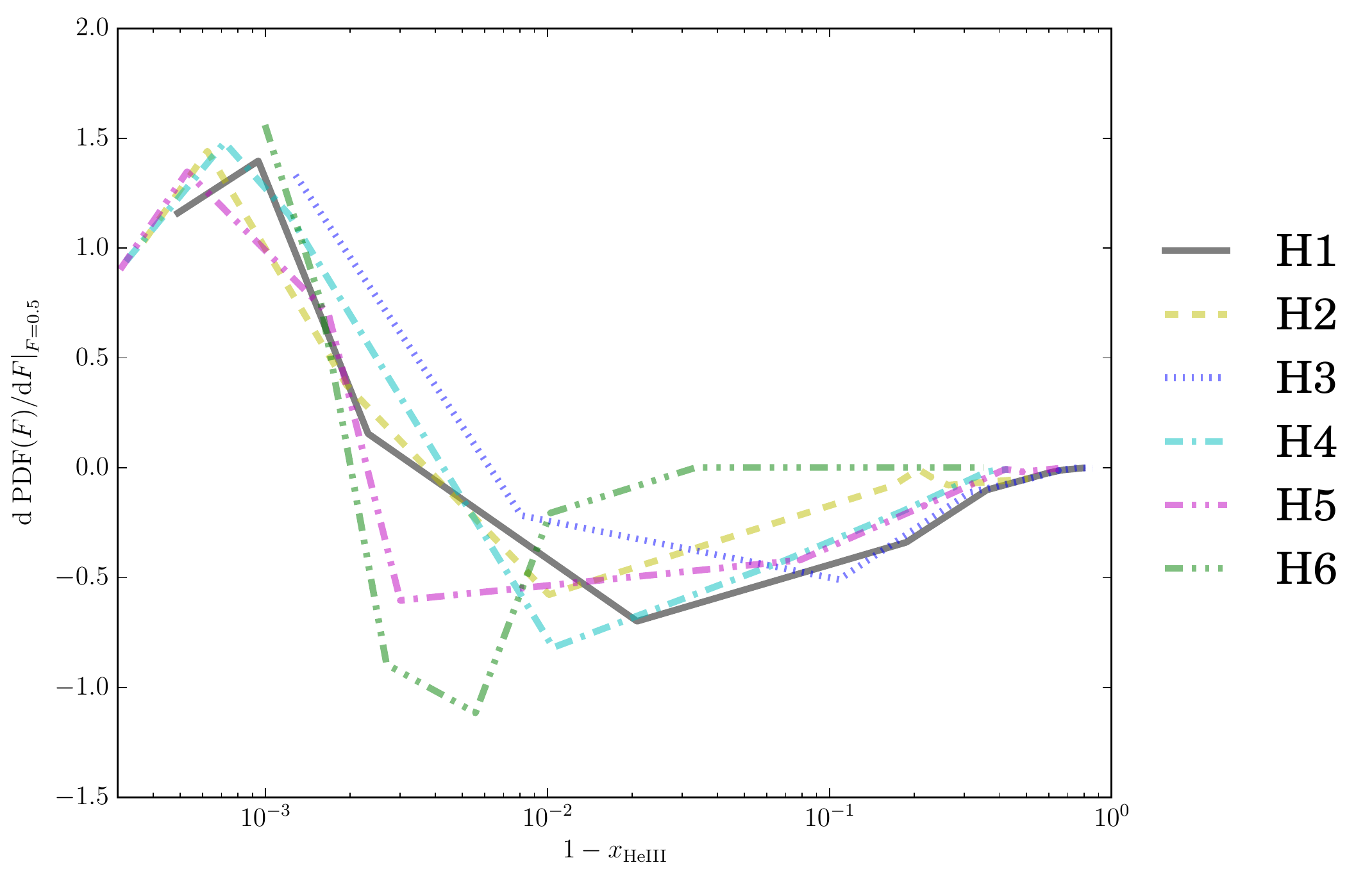}
  \caption{The derivative of the flux PDF at $F = 0.5$ as a function of the
    helium~\textsc{iii} fraction for the different simulations. The complement
    of the helium~\textsc{iii} fraction $1 - x_\mathrm{HeIII}$ is shown, to
    emphasize the behavior at high ionization fractions. The helium fraction is
    shown instead of redshift to emphasize characteristics common to the
    different reionization scenarios. At early times (when
    $1 - x_\mathrm{HeIII} \sim 1$), the slope of the flux PDF is typically flat
    or negative. The slope is negative for the majority of the ionization
    process, but becomes positive again following a 99\% ionization
    fraction. This feature is common across quasar models, and thus represents a
    robust indicator of the timing for the end of helium reionization. Note that
    the simulations H3 and H6 do not reach ionization fractions greater than
    99.9\% ionized by $z \sim 2$.}
  \label{fig:xHeII_pdf}
\end{figure}

Figure~\ref{fig:fluxPDF} shows the \HeII\ flux PDF for the different
simulations. The panels show the volume at redshifts $z \sim 2.7$ (top),
$z \sim 2.5$ (bottom left), and $z \sim 2.3$ (bottom right). Note that at
redshift $z \sim 2.7$, most of the simulations are 99\% reionized. Despite this
fact, there are comparatively few pixels with high transmission: for the
fiducial case of Simulation~H1, 90\% of the pixels have flux of $F \leq
0.5$. This relatively strong absorption is related to the strength of the
Ly$\alpha$ transition, where only a small amount of \HeII\ is necessary to
absorb most of the incoming radiation. Note, though, that measurement of the
flux PDF can still be an important marker of the timing of reionization. As
noted above, Simulation~H3 reaches 99\% reionization significantly later at
$z_{99} \sim 2.23$, which is evident in the distinct shape of the flux PDF. In
particular, there are far fewer pixels with $F \geq 0.5$ at all redshifts,
indicative of its relatively late completion of reionization. The flux PDF of H3
at $z \sim 2.3$ is comparable to that of, \textit{e.g.}, Simulation~H1 at
$z \sim 2.7$. Thus, by measuring the redshift when the central portion of the
flux PDF is relatively flat (\textit{e.g.}, when
$PDF(F=0.25) \approx PDF(F=0.75)$), one can determine the timing of when the
volume is $\sim$99.9\% ionized. This point is addressed more directly in
Figure~\ref{fig:xHeII_pdf} (discussed below).

\begin{figure*}[t]
  \centering
  \resizebox{0.6\textwidth}{!}{\includegraphics{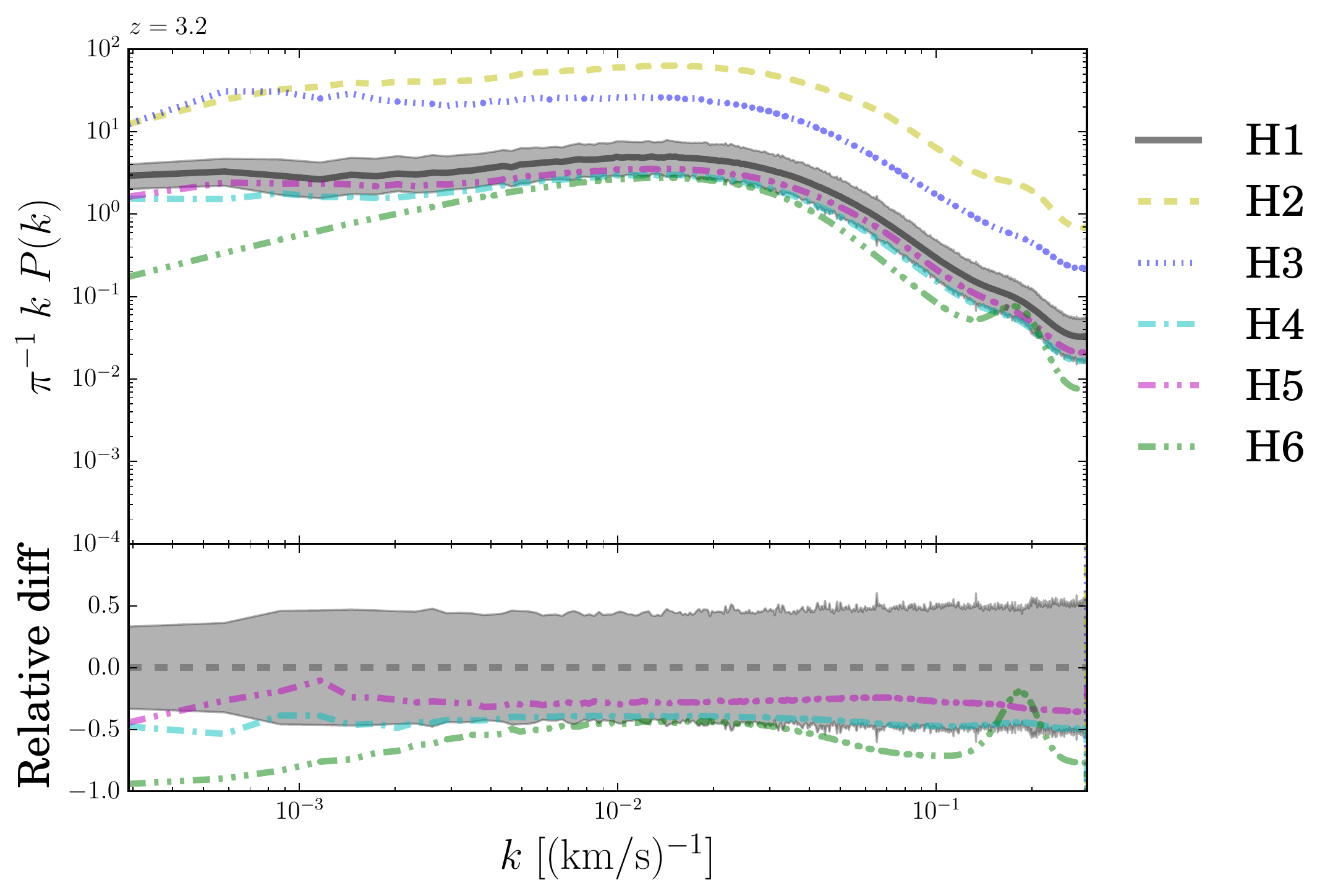}}\\
  \resizebox{0.48\textwidth}{!}{\includegraphics{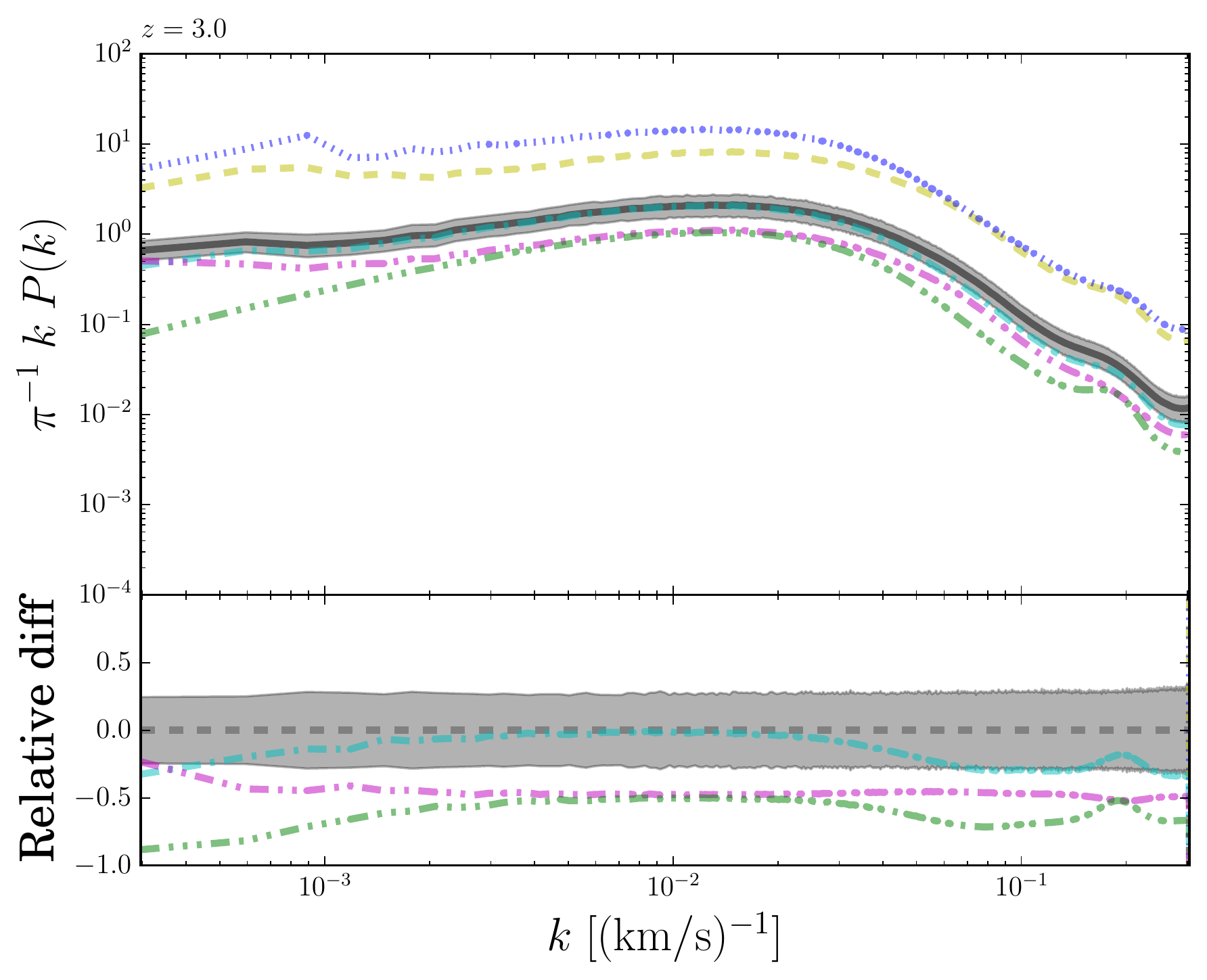}}%
  \resizebox{0.48\textwidth}{!}{\includegraphics{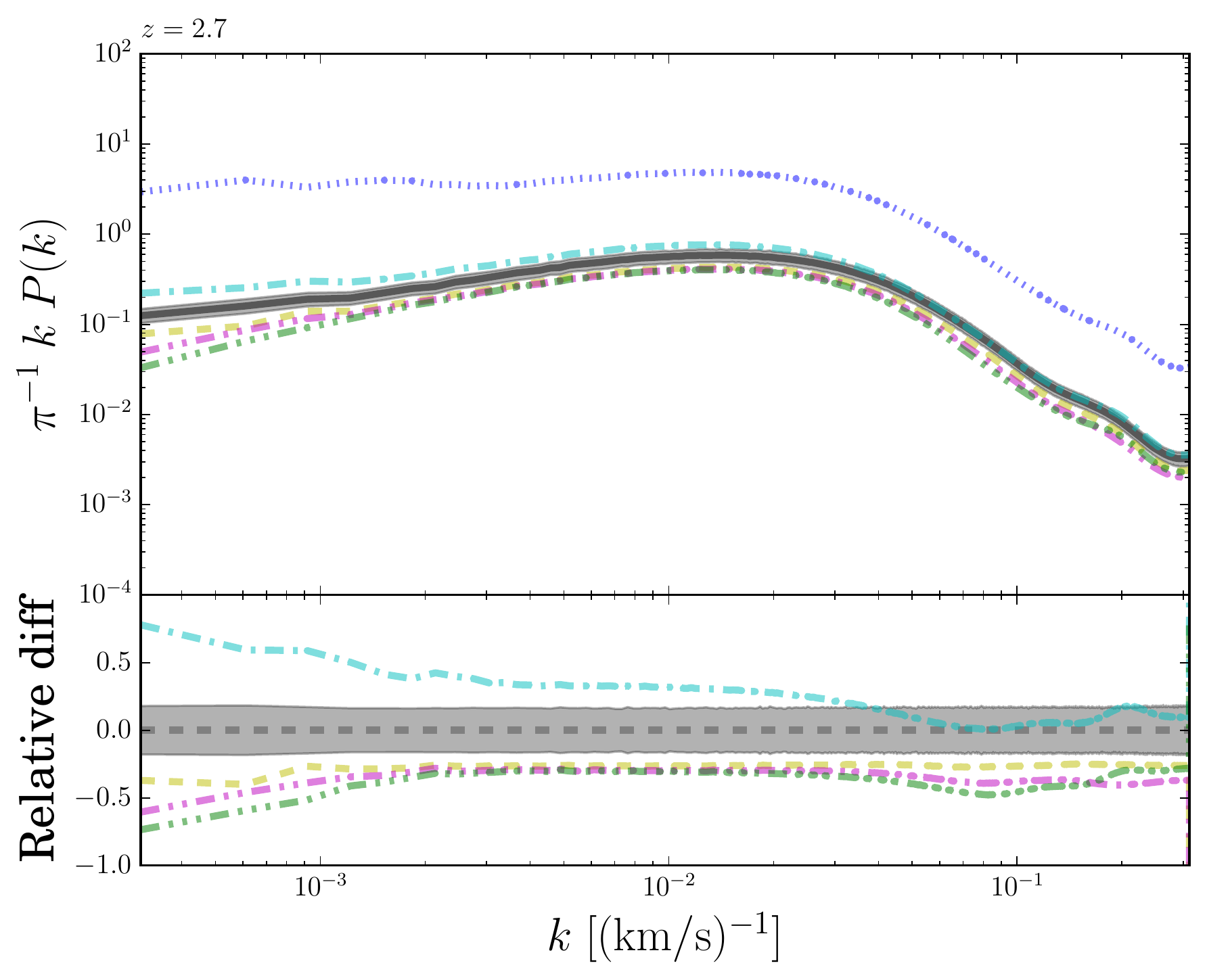}}\\
  \caption{One-dimensional flux power spectra of the \HeII\ Ly$\alpha$ forest,
    at redshift $z \sim 3.2$ (top), $z \sim 3$ (bottom left), and $z \sim 2.7$
    (bottom right). These redshifts are different than the ones in
    Figure~\ref{fig:fluxPDF} because the change in amplitude of the power
    spectra are more evident earlier in the reionization process. At a given
    redshift, there is a marked difference in the overall amplitudes of the
    power spectrum. These changes are correlated with the value of
    $\tau_\mathrm{HeII}$ at a given redshift. Compare to
    Figure~\ref{fig:tauHeII}, and note that the amplitude of the power spectrum
    largely tracks the values of $\tau_\mathrm{eff}$. See the text for
    additional discussion.}
  \label{fig:1dps}
\end{figure*}

The shaded error regions in Figure~\ref{fig:fluxPDF} show 1$\sigma$
uncertainties, and are calculated using bootstrap resampling of 50 sightlines
and computing the variance within each flux bin.\footnote{Note though that, as
  mentioned in Section~\ref{sec:taueff}, using 50 samples in the bootstrap
  realization for determining the variance may be overly optimistic.} Moreover,
the sightline length used is 10 proper Mpc, which is comparable to the distances
reported in measurements \citep{worseck_etal2014}. In contrast to the shaded
regions in Figure~\ref{fig:tauHeII}, these error regions are generally quite
small, and do not significantly overlap with other simulations. This result
demonstrates that only a few sightlines are necessary to determine the shape of
the flux PDF. It should be noted that analogously to the \HI\ flux PDF
(and further discussed in Appendix~B of Paper~II), the continuum level of the
measured spectra can have a dramatic effect on the shape of the flux
PDF. Accordingly, uncertainty in this level can lead to systematic shifts in the
calculated flux PDF. Nevertheless, Figure~\ref{fig:fluxPDF} shows that given low
systematic uncertainties, the flux PDF of the \HeII\ Ly$\alpha$ forest can
be a powerful tool for determining the ionization state of helium in the IGM.

Figure~\ref{fig:xHeII_pdf} shows the derivative of the flux PDF
$\dv*{\ \mathrm{PDF}(F)}{F}$ at a value of $F=0.5$ as a function of the
complement of \HeIII\ ionization fraction $1 - x_\mathrm{HeIII}$. This quantity
is used instead of simply the ionization fraction to emphasize the behavior at
high ionization levels. All of the reionization scenarios have been converted
from redshift to ionization fraction in order to demonstrate uniformity across
realizations. As discussed above, the transition from most pixels with values of
$F \sim 0$ to $F \sim 1$ occurs relatively late in the ionization process. This
transition can be captured in the change of the slope of the PDF at a value of
$F = 0.5$: at early times and low ionization levels, the slope is negative with
increasing flux values. Once the volume becomes $\sim$99\% ionized, the flux PDF
flattens out. At even higher ionization levels, the slope becomes positive. As
shown in Figure~\ref{fig:xHeII_pdf}, the timing for these transitions is closely
tied to the ionization fraction, rather than a specific redshift. Furthermore,
as demonstrated with the relatively small statistical error bars in
Figure~\ref{fig:fluxPDF}, the flux PDF can be determined to high fidelity with
relatively few \HeII\ sightlines. Assuming systematic uncertainties of
observations can be sufficiently mitigated, this quantity represents a robust
method for determining the endpoint of helium~\textsc{ii} reionization, to a
much greater degree than the effective optical depth $\tau_\mathrm{eff,HeII}$.

\subsection{One-dimensional flux power spectra}
\label{sec:1dps}
In addition to the flux PDF, the one-dimensional power spectrum of the \HeII\
Ly$\alpha$ forest can be used to learn important information about the
ionization state of the IGM. The overall amplitude of the power spectrum as well
as the shape as a function of Fourier mode $k$ will change as the ionization
state and the size of ionized regions change. As with the one-dimensional power
spectrum for \HI, the amplitude on large scales is related to the optical depth
$\tau_\mathrm{eff,HeII}$, with higher amplitudes corresponding to higher values
of $\tau_\mathrm{eff,HeII}$ (see Figure~\ref{fig:tauHeII}).

Figure~\ref{fig:1dps} shows the one-dimensional power spectrum of the \HeII\
Ly$\alpha$ forest. The primary difference between the simulations is in the
amplitude of the power spectra. At a given redshift, the amplitude of the power
spectrum is directly related to the value of $\tau_\mathrm{eff}$. Note that
Simulation~H3 has the largest value of $\tau_\mathrm{eff}$ at a given redshift
(as shown in Figure~\ref{fig:tauHeII}), and also has the largest amplitude in
Figure~\ref{fig:1dps}. This can be understood in terms of the amplitude of
fluctuations in the flux field: when the IGM has a relatively low value of
$\tau_\mathrm{eff}$, then all points in the volume have a similarly (high) value
of flux. These differences in flux are driven primarily by correlations with the
radiation field. Not only is the fraction of \HeIII\ higher in regions of high
radiation intensity, but the temperature is also greater. Both of these effects
contribute to a lower value of $\tau_\mathrm{eff,HeII}$, or a higher value of
flux $F$. The combination leads to highly correlated regions of high flux and
low flux, increasing the amplitude of the power spectrum. At the same time, it
is not solely the radiation field that controls the amplitude of the power
spectrum, as Simulation~H6 has a uniform radiation field. In this simulation,
the differences are driven primarily by the local gas density, and so there is
no corresponding correlation between regions of high and low flux.

At $z \sim 2.7$, the difference in the power spectrum amplitude of Simulation~H3
at all scales is an order of magnitude larger than that of the other
simulations. Such a dramatic difference should be detectable, and would allow
for a straightforward determination of the helium ionization state of the
IGM. Most importantly, the amplitude of the flux power spectrum as a function of
redshift are clear and pronounced, even for reionization histories that are not
fully reionized. Hence, the one-dimensional power spectrum can be a window into
helium~\textsc{ii} reionization at times prior to 99\% ionization.

The shaded regions show 1$\sigma$ uncertainty in the measurements using 50
sightlines and bootstrap resampling, with the same rationale as that discussed
regarding Figure~\ref{fig:fluxPDF} in Section~\ref{sec:fluxpdf}. Unlike the
approach taken there, though, the entire sightline is used rather than a 10 Mpc
segment. By using the entire sightline, there are no issues related to broken
periodicity when performing the Fourier transform for the power spectrum
calculation.\footnote{Note though that at $z \sim 3$, the sightlines from the
  simulation are $\sim 71$ proper Mpc, which is almost an order of magnitude
  larger than the observational sightlines (which as stated above are $\sim 10$
  proper Mpc). There is thus implicitly additional information in each of the
  simulated sightlines compared to the observational ones, though there are
  fewer numerical artifacts introduced by using the whole sightline and not
  explicitly breaking periodicity in the Fourier transform.} At the earliest
redshift ($z \sim 3.2$), there is a relatively large uncertainty, so that at
most scales, several of the reionization histories are expected to lie within
1$\sigma$ of each other. However, simulations with vastly different values of
$\tau_\mathrm{eff}$ (as in Simulations~H2 and H3 in Figure~\ref{fig:tauHeII})
still show a distinct change in amplitude of the power spectrum. Thus, the
one-dimensional power spectrum can serve as another measurement of the overall
opacity of the volume.

At lower redshift, the uncertainty of the power spectra decreases noticeably. As
a result, in principle it becomes easier to distinguish the histories. At the
same time, there is significant overlap in several of the histories, which is
due to having comparable value of $\tau_\mathrm{eff}$. Some of the largest
differences that remain are at large scales. As with the \HI\ Ly$\alpha$ forest
(and discussed in Paper~II), these differences might be attributable to the
large-scale radiation field. Because the radiation field is highly non-uniform
\HeII -ionizing radiation originating from quasars (as opposed to the \HI\
forest that has a largely uniform background component from galactic radiation),
the large-scale power may reflect the degree of bias in the sources. Note in
particular that Simulation~H6, which has only a uniform UV background and no
explicit sources, has consistently the lowest large-scale power, despite having
one of the earliest reionization times. At all redshifts considered, this
simulation shows a lack of power compared to the simulations with explicit
sources. Thus, the large-scale power may be a way to learn about the bais of
sources of helium~\textsc{ii} reionization.

Nevertheless, there are potential observational complications associated with
determining the one-dimensional flux power spectrum. Due to the overall low
value of flux in the \HeII\ Ly$\alpha$ forest before the conclusion of
reionization, it is difficult to determine the continuum level, and hence the
flux measurement. Determining the global (low) flux value can lead to an
incorrect normalization, and hence raise or lower the large-scale power spectrum
amplitude. Despite this difficulty, the one-dimensional power spectrum remains
one of the few probes that can detect the ionization level of helium before
completion, and does not rely on calibrating other measurements of the IGM.

\section{Discussion}
\label{sec:discussion}
One very pertinent question with these measurements is the degree to which the
reionization history can be determined with a limited number of observations. As
discussed in Section~\ref{sec:intro}, to date there have been only about 50
observations of the \HeII\ Ly$\alpha$ forest \citep{syphers_etal2012}. We
have shown in Secs.~\ref{sec:fluxpdf} and \ref{sec:1dps} that the flux PDF and
one-dimensional power spectrum provide significant information about the
ionization state of helium. However, it is reasonable to wonder to what extent
current observations are able to determine the ionization state of the IGM.

To this end, we have used bootstrap resampling using 50 sightlines to estimate
the standard deviation for our different scenarios. Figures~\ref{fig:fluxPDF}
and \ref{fig:1dps} show the 1$\sigma$ dispersion as measured for 50
sightlines. As noted in the earlier discussion, the ionization state of helium
may be readily detectable in the flux PDF measurement. Even accounting for
uncertainty in the continuum level of the forest, the shape of the flux PDF
varies strongly as a function of ionization fraction. This variation in shape is
significantly larger than the inherent variation of the flux PDF, and so even
with comparatively few sightlines, a meaningful determination of the ionization
level of helium maybe possible given the current
data. Figure~\ref{fig:xHeII_pdf} shows that the slope of the flux PDF is capable
of well-characterizing the timing of the end of reionization, since the slope is
highly correlated with the ionization fraction rather than a particular
redshift. Accordingly, the flux PDF should be a powerful tool for learning more
about helium~\textsc{ii} reionization.

Furthermore, the \HeII\ one-dimensional power spectrum on large scales could
help determine the bias of sources driving helium~\textsc{ii} reionization. Note
that there is more than an order of magnitude difference in the amplitudes at
large scales (\textit{e.g.}, $k = 3\times 10^{-3}$ (km/s)$^{-1}$), which should
be detectable. In addition to the overall amplitude, the shape of the power
spectrum on large scales is greatly influenced by the relative bias of the
sources: note that all simulations with explicit sources (H1-H5) have a
relatively flat power spectrum at large scales, but the uniform UV background
featured in Simulation~H6 shows decreasing amplitude with decreasing $k$. This
is likely related to the radiation field properties: the differences in the
helium~\textsc{ii} ionization fraction in Simulation~H6 are driven primarily by
gas density fluctuations, since the same ionization field is seen at all points
in the volume. Accordingly, there is less correlation between regions of high
flux and low flux in terms of helium ionization level as well as temperature,
and thus less power. At the same time, determining the continuum level in
observations is difficult for such small flux levels, and errors in its
determination may power spectrum amplitude.

\section{Conclusion}
\label{sec:conclusion}
To date, the \HeII\ Ly$\alpha$ forest has largely only been used to
determine the value of $\tau_\mathrm{eff,HeII}$. As can be seen from
Figure~\ref{fig:tauHeII}, there is a very large dispersion in this measurement,
owing to the large sightline-to-sightline variations. Thus, determining the
reionization history from this quantity alone is very difficult, and leads to
large uncertainties in the determination of the redshift of
reionization. Additionally, as discussed in Section~\ref{sec:taueff}, this
measurement is largely sensitive to the tail-end of reionization, and does not
yield much information about the intermediate stages of the reionization
process. Accordingly, new applications of the \HeII\ Ly$\alpha$ forest would
be beneficial for learning more about the timing and duration of reionization.

To this end, we have presented the flux PDF and the one-dimensional power
spectrum as ways to break the degeneracy present in $\tau_\mathrm{eff}$. These
differences are generally quite large between different simulations, in some
cases being larger than an order of magnitude. Further surveys will hopefully be
able to take advantage of these pronounced differences, and begin to measure the
timing and duration of helium~\textsc{ii} reionization.

\acknowledgements{This work was supported in part by NASA grants NNX14AB57G and
  NNX12AF91G and NSF grant AST15-15389.}

\bibliography{mybib}
\bibliographystyle{apj}

\end{document}